\documentclass[aps,prb,amsmath,amssymb,reprint,superscriptaddress,longbibliography]{revtex4-2}

\usepackage{dcolumn}
\usepackage{bm}
\usepackage{xcolor}
\usepackage{xr-hyper}
\usepackage{amsmath,amsfonts,amssymb}
\usepackage{graphicx}
\usepackage{siunitx}
\usepackage{natmove}
\usepackage{hyperref}
\usepackage{multirow}
\usepackage{array}

\newcommand{\mycomment}[1]{}
\newcommand{\edit}[1]{#1}

\newcommand{\norm}[1]{\left|{#1}\right|}

\newcommand{\panel}[1]{\textbf{(#1)}~}

\makeatletter
\renewcommand{\subsection}{%
  \@startsection
    {subsection}
    {2}
    {0pt}
    {0.5\baselineskip plus 0.5\baselineskip minus 0.5\baselineskip}
    {0.5\baselineskip plus 0.5\baselineskip}
    {\small\bfseries\centering}
}
\makeatother

\begin{document}
%
%
\title{Loop-gap resonators achieving strong magnon--photon coupling in magnetic insulator thin films}
%

\author{Francesca Zanichelli}
\thanks{These authors contributed equally}
\affiliation{Department of Materials, ETH Zurich, H{\"{o}}nggerbergring 64, 8093 Zurich, Switzerland}
\affiliation{Department of Physics, {\'{E}}cole Polytechnique, Institut Polytechnique de Paris, 91128 Palaiseau, France}
\author{Davit Petrosyan}
\thanks{These authors contributed equally}
\author{Hanchen Wang}
\author{Patrick Helbingk}
\affiliation{Department of Materials, ETH Zurich, H{\"{o}}nggerbergring 64, 8093 Zurich, Switzerland}
\author{Richard Schlitz}
\affiliation{Department of Materials, ETH Zurich, H{\"{o}}nggerbergring 64, 8093 Zurich, Switzerland}
\affiliation{Department of Physics, University of Konstanz, 78457 Konstanz, Germany}
\author{Pietro Gambardella}
\affiliation{Department of Materials, ETH Zurich, H{\"{o}}nggerbergring 64, 8093 Zurich, Switzerland}
\author{William Legrand}
\email{william.legrand@neel.cnrs.fr}
\affiliation{Department of Materials, ETH Zurich, H{\"{o}}nggerbergring 64, 8093 Zurich, Switzerland}
\affiliation{Universit{\'{e}} Grenoble Alpes, CNRS, Institut N{\'{e}}el, 38042 Grenoble, France}

%
%
%
\begin{abstract}
Magnon--photon hybrid systems consisting of a three-dimensional electromagnetic resonator and a bulk magnetic insulator constitute the standard experimental platform in cavity magnonics. Here, we demonstrate a modular loop-gap resonator design optimized to couple with thin films of magnetic insulators. We achieve the strong-coupling regime using this loop-gap resonator coupled to a 75~nm-thick epitaxial film of yttrium iron garnet at room temperature. We further show how to perform field-differential spectroscopy of the hybrid magnon--photon system, which eliminates the unwanted signal from other loop-gap modes uncoupled to the magnetic film. In addition to the uniform ferromagnetic resonance mode, the loop-gap resonator enables an hybridization with the standing spin-wave modes forming across the thickness of the film. Our approach unlocks the use of epitaxial films and multilayers of magnetic insulators to tune the magnon band structure in cavity magnonics experiments.
\end{abstract}
\maketitle
%

\section{Introduction}

Magnon--photon hybrid systems offer a promising platform for investigating the magnon dynamics of exchange-coupled spin ensembles \cite{Huebl2013,Rameshti2022}. Previous studies have examined how hybridized magnons and microwave photons interact with spin currents \cite{Bai2015,MaierFlaig2016}, their coupling with optical light and whispering gallery modes \cite{Hisatomi2016,Haigh2016,Osada2016}, and how their hybridization changes in the presence of dissipative couplings \cite{Harder2018a,Zhang2019}. The magnetic part in these systems can  be an antiferromagnet \cite{Everts2020,Boventer2023}, they feature the magnon Kerr effect \cite{Wang2016,Wang2018,Lee2023}, and are able to couple coherently with phonons to form cavity magneto-mechanical systems \cite{Zhang2016}.

Experiments in the field of cavity magnonics typically rely on hollow cavities or three-dimensional lumped-element resonators \cite{Goryachev2014,Flower2019a,Potts2020}, which are most suitable for coupling with bulk crystals of magnetic insulators. A simplified figure of merit for hybrid systems is the cooperativity $\mathcal{C}=4g^2/(\kappa_{\rm{c}} \kappa_{\rm{m}})$, exemplifying the ratio between the magnon--photon coupling strength ($g$) and the photonic ($\kappa_{\rm{c}}$) and magnonic ($\kappa_{\rm{m}}$) dissipation rates \cite{Rameshti2022}. Coherent hybridized modes require the strong-coupling regime $g> \kappa_{\rm{c}},\kappa_{\rm{m}}$. In a given resonator design, $g$ increases with the number of spins $N$ participating to the collective magnon modes as $\sqrt{N}$. For this reason, ultrathin ($<$~\SI{100}{\nano\meter}) films of magnetic insulators have so far remained impractical for use in magnon--photon hybrid systems, considering their reduced size compared to bulk crystals.

Single-crystal spheres of yttrium iron garnet (YIG), commonly used in cavity magnonics, feature record-low magnetic resonance linewidths. They also exhibit an almost continuous succession of overlapping and intersecting magnetostatic modes \cite{Walker1958,Fletcher1959,Gloppe2019}, which prevents the isolation of a single magnon mode. The decay of the principal magnon mode into spatially-dependent modes is an essential limit to the coherence of the magnon states \cite{Lee2023}. By contrast, thin films of magnetic insulators exhibit well separated modes and considerably improved spectral distance between them. The magnetostatic anisotropy of thin films, as opposed to the isotropic shape of spheres, introduces an efficient means of structuring the magnon energy spectrum and magnon--magnon couplings, which is complemented by the possibility to further tune the anisotropy by growth and strain engineering \cite{Wang2014b,Soumah2018}. Moreover, epitaxial films of magnetic insulators can feature compositions not easily stabilized in bulk \cite{Legrand2025a} and can be grown as heterogeneous multilayers \cite{Khurana2024}, unlocking new magnon properties. Replacing bulk crystals with thin films of magnetic insulators would therefore bring considerable advantages in cavity magnonics.

The cooperativity of magnon--photon systems comprising a magnetic insulator thin film can be enhanced by increasing the filling factor of the resonator, i.e., by adapting its geometry to maximize the ratio between the volume of the magnetic sample and the confined volume of the electromagnetic field in the resonator. Accordingly, prior works have employed low-temperature planar integrated resonators, made from superconducting materials to avoid the resistive losses originating from miniaturization \cite{Huebl2013,Li2019,Hou2019,McKenzie-Sell2019}. Unless nanomagnets are used, most planar integrated resonators provide an inhomogeneous electromagnetic field, which can then excite not only the main uniform magnon resonance, but also all modes with finite wave-vectors. Between hollow cavities and integrated planar resonators \cite{Akhmetzyanov2023}, lumped-element resonators represent an intermediate solution that can operate at both low and room temperatures. These resonators have previously been applied to solid-state paramagnetic spin ensembles \cite{Angerer2016,Ball2018,Choi2023}. Notably, loop-gap geometries are commonly used in electron spin resonance to provide a homogeneous microwave field distribution throughout the sample and maximize the filling factor of compact resonators, across a wide range of frequencies \cite{Simovic2006,Sidabras2007}.

\begin{figure*}[t]
    \centering
    \includegraphics[width=\textwidth]{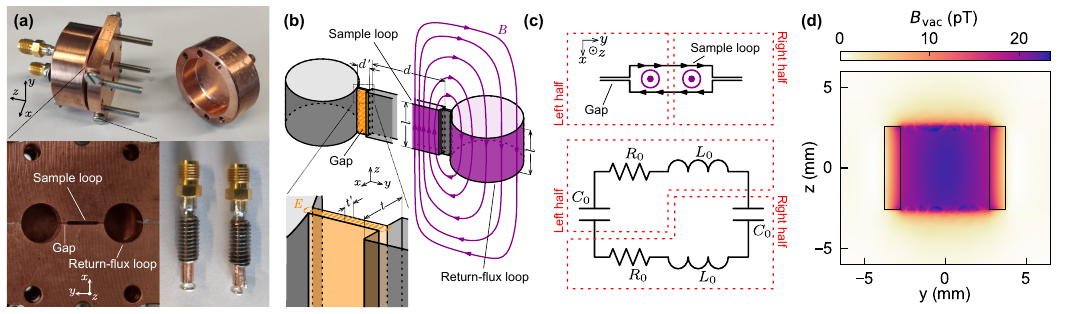}
    \caption{\panel{a}Photographs of the resonator system: assembled LGR and top half of cylindrical box next to bottom half of cylindrical box; detail of the loops and gaps; coupling antennas with screw system. \panel{b}Schematic representation of the LGR designed in this work: the $E$-field (orange) is mostly confined in the gaps, as shown in zoomed-in view; the $B$-field (purple) is concentrated in the sample loop and is residual in the return-flux loops used for external coupling. \panel{c}Sketch of current lines, and equivalent circuit model of the LGR as symmetrical lumped $R_0$, $L_0$ and $C_0$. \panel{d}Intensity map of the coupling field $B\textsubscript{vac}$ in the sample plane ($yz$), as determined by finite-element simulations. The black rectangles locate the two gaps.}
    \label{fig:01}
\end{figure*}

In this work, we demonstrate a modular loop-gap resonator (LGR) optimized to couple with thin films of magnetic insulators as a platform for cavity magnonics. The LGR structure focuses the magnetic component of the electromagnetic field to a small mode volume matching the sample's dimensions, and enables homogeneous coupling to the magnetic system. We use finite-element modeling to optimize the LGR geometry and its probing. We assess the LGR performance by microwave measurements of the resonator loaded with a \edit{\SI{75}{\nano\meter}-thick} YIG thin film, which reaches $\mathcal{C}$ above unity and the strong-coupling regime. We also show how to perform field-differential magneto-spectroscopy \cite{Legrand2025} of this hybrid system, which isolates the mode of interest (coupled to magnons) relative to the other modes of the resonator (decoupled from magnons \cite{Goryachev2014}). The sensitivity of the probing scheme enabled by  the LGR also allows for detecting perpendicular standing spin-wave modes with very weak surface pinning, despite their low external susceptibility to homogeneous probing fields. These results and the modularity of the LGR design open up the investigation of epitaxial magnetic thin films in various cavity magnonics experiments and across a broad temperature range.


\section{Resonator design}

The magnetic resonance linewidth of epitaxial magnetic insulator thin films is typically in the range $\kappa_{\rm{m}}/(2\pi)\approx$~10--\SI{100}{\mega\hertz}. Since resistive losses in microwave conductors increase when reducing their dimensions, an LGR constitutes an advantageous trade-off between mode volume and resistive losses, with a room-temperature dissipation $\kappa_{\rm{c}}/(2\pi)$ comparable to that of the magnetic system. Its reduced mode volume compared to hollow resonators provides stronger coupling, sufficient even for films with \si{\nano\meter} thickness. By contrast, a monolithic integrated approach such as an on-chip coplanar waveguide, featuring much narrower conductors, is not suitable at room temperature due to excessive resistive losses. Alternative approaches to further reduce the mode volume by exploiting the sub-wavelength confinement of the electromagnetic field in a resonator have been reported \cite{Choi2023}, but come at the expense of magnified resistive losses and a strongly inhomogeneous probing field. The inhomogeneous probing field can then couple to multiple magnon modes with finite wave-vectors, which may often partially overlap. For a thin film with \si{\milli\meter} lateral dimensions, the filling factor in the magnetic mode volume must be as large as possible. In such a case, the gain in single-spin coupling obtained by reducing further the mode volume will be directly compensated by the reduced volume of the magnetic system and reduced $N$. Therefore, we focus on designs that maximize the filling factor, in order to approach the optimal cooperativity for a fixed spin number density $n_{\rm{s}}$ (for instance, \SI{2e28}{\per\meter\cubed} in YIG) of the magnetically ordered medium.

\subsection{LGR properties}
Our resonator (Fig.~\ref{fig:01}a) is composed of two oxygen-free high-conductivity (OFHC) copper parts, each forming half of an LGR. The assembled module comprises two narrow slits (gaps), an opening for the central sample space (sample loop), and two additional loops which serve as a return path for the magnetic flux (return-flux loops). The two halves are assembled by indium bonding on their conductive surfaces, and enclosed in a cavity box with probing antennas. The external coupling of the LGR can be continuously and precisely adjusted by tuning the position of the probing antennas along $z$ using a screw system (see Fig.~\ref{fig:01}a). Due to the lumped-element geometry, the magnetic component of the electromagnetic field (purples lines in Fig.~\ref{fig:01}b) is efficiently concentrated in the sample space, and the electrical component (orange lines in Fig.~\ref{fig:01}b) is concentrated in the gaps.

\footnotesize
\begin{table*}[t]
    \centering
    \caption{Characteristic parameters for the final LGR1 design: $l=\SI{5.1}{mm}$, $d=\SI{5.5}{mm}$, \edit{$t =\SI{0.55}{mm}$}, $d'=\SI{1.0}{mm}$, \edit{$t'=\SI{0.1}{mm}$}, $S_{\rm{L}'}=\SI{39}{mm^2}$ for each return loop. Columns are model type: $R,L,C$ or finite-element model (FEM) with vacuum or different substrates inserted: GGG ($\varepsilon_{\rm{r}}=12$), YSGG ($\varepsilon_{\rm{r}}=13$) or YIG ($\varepsilon_{\rm{r}}=16$); resonance frequency; quality factor with/without sample inserted but no external coupling from antennas; coupling magnetic field expressed as $B_{\rm{vac}}^{\rm{max}}$, its value at the LGR center; corresponding $B_{\rm{vac}}^{\rm{rms}}$ averaged over the sample volume; resonance linewidth; filling factors $\eta$ of the sample loop of \qtyproduct[product-units=power]{5.5 x 5.1 x 0.55}{\milli\metre} and $\eta_{\rm{S}}$ of a sample of \qtyproduct[product-units=power]{5 x 5 x 0.5}{\milli\metre} within the magnetic mode volume; and standard deviation of $B_{\rm{vac}}^{\rm{rms}}$ in the \qtyproduct[product-units=power]{5 x 5 x 0.5}{\milli\metre} sample volume. The $R,L,C$ model cannot estimate the filling factor and standard deviation of $B_{\rm{vac}}^{\rm{rms}}$.}
    \label{tab:sim_results}
    
    \begin{tabular*}{\textwidth}{@{\extracolsep{\fill}}lcccccccc}
        \hline\hline
        & $\omega_{\rm{0}}/(2\pi)$ 
        & \multirow{2}{*}{$Q^\mathrm{unloaded}$} 
        & $B_{\rm{vac}}^{\rm{max}}$
        & $B_{\rm{vac}}^{\rm{rms}}$
        & $\kappa_{\rm{c}}/(2\pi)$ 
        & \multirow{2}{*}{$\eta$} & \multirow{2}{*}{$\eta_{\rm{S}}$} & \multirow{2}{*}{$\sigma_{\rm{B}}/B$} \\
        & (GHz) & & (pT) & (pT) & (MHz) & & \\
        \hline
        $R,L,C$ model & 12.27 & 720 & - & 12.9 & 17.0 & - & - & - \\
        FEM simulation, without sample & 11.07 & 851 & 22.6 & 15.0 & 13.0 & 0.753 & 0.620 & 0.054 \\
        FEM simulation, GGG substrate & 9.15 & 831 & 25.1 & 13.4 & 11.0 & 0.777 & 0.674 & 0.246 \\
        FEM simulation, YSGG substrate & 9.02 & 831 & 25.0 & 13.3 & 10.9 & 0.777 & 0.674 & 0.248 \\
        FEM simulation, YIG substrate & 8.91 & 827 & 25.3 & 13.2 & 10.8 & 0.780 & 0.679 & 0.268 \\
        \hline\hline
    \end{tabular*}

\end{table*}
\normalsize

Figure~\ref{fig:01}c presents an equivalent circuit model of this LGR \cite{Rinard1993, Rinard2005, Wood1984}, in which $R$, $L$ and $C$ elements coupled in series link the resonance frequency, linewidth, and magnon--photon coupling strength to the geometrical dimensions of the design. Here, $l$ is the length of the resonator along $z$; $d$ and $d'$ are the sample loop width and gap width along $y$; $t$ and $t'$ are \edit{the sample loop height and the gap separation} along $x$, respectively. Each gap corresponds to a parallel-plate capacitor \edit{$C_0=\varepsilon_0\varepsilon_{\rm{r}} S_{\rm{G}}/t'$} with $\varepsilon_0$ the vacuum dielectric permittivity, $\varepsilon_{\rm{r}}\approx1$ the relative permittivity of the air-filled gap, $S_{\rm{G}}=ld'$ the plate surface. Each side of the sample loop is half of a single-turn solenoid, of cross-section \edit{$S_{\rm{L}}/2=dt/2$}, and has self-inductance $L_0=\mu_0 S_{\rm{L}}/2l$ with $\mu_0$ the vacuum magnetic permeability. The resistors are essentially the sample loop walls, and \edit{$R_0=R_{\rm{s}}(d+t+d')/l$}, where $R_{\rm{s}}=\sqrt{\omega_0\mu_0/(2\sigma_{\text{Cu}})}$ is the surface resistivity and $\sigma_{\text{Cu}}$ the conductivity of OFHC copper. In series, the lumped-element parameters are $C=C_0/2$, $L=2L_0$ and $R=2R_0$, and the resonance frequency is $\omega_0/(2\pi)=1/(2\pi\sqrt{LC})$ with unloaded quality factor $Q=\omega_0 L/R$ \cite{Pozar2012}. 

The single-spin coupling strength is given by $g_0/(2\pi)=(\gamma/2\pi)B_{\rm{vac}}^{\rm{rms}}\langle{-}|\mathbf{S}_x|{+}\rangle$ with $\mathbf{S}_x$ the spin operator orthogonal to the static field and $|{+}\rangle,|{-}\rangle$ the spin eigenstates along the static field, leading to $g_0/(2\pi)=(\gamma/2\pi)B_{\rm{vac}}^{\rm{rms}}/2$, where $\gamma$ is the electron spin gyromagnetic ratio and $B_{\rm{vac}}^{\rm{rms}}$ is the rms magnetic component of the vacuum fluctuations in the LGR. Approximating the magnetic component of the electromagnetic field in the LGR to be homogeneous (this is a property of the LGR geometry), $B_{\rm{vac}}^{\rm{rms}}=\sqrt{\mu_0\hbar\omega_0/(2V_{\rm{m}})}$ applies to the whole magnetic system, with $V_{\rm{m}}$ the mode volume of the LGR. Under the simplifying assumption that the entire magnetic energy is confined in the sample loop (see Fig.\ \ref{fig:01}d and below), $V_{\rm{m}}$ can be replaced by \edit{$V=lS_{\rm{L}}=ldt$}. These simple relations provide useful scaling laws for a first optimization.

The choices leading to our final design are detailed in Appendix~\ref{app:optim}. To evaluate the LGR properties, we rely on a finite-element electromagnetic solver (Ansys HFSS), which accounts for the imperfect separation of $E$- and $B$-fields that causes $C$ to vary with $\varepsilon_{\rm{r}}$ of the sample, and for deviations from ideal lumped-element behavior in the actual surface current distributions. A comparison between finite-element modeling and the simplified circuit model above is provided in Table~\ref{tab:sim_results}. The LGR design is adjusted to reach a resonant frequency $\omega_0/(2\pi)\approx\SI{10}{\giga\hertz}$. The uncoupled LGR quality factor is predicted to be $Q=851$, mostly limited by the unavoidable surface resistivity of OFHC copper at room temperature, with $\sigma_{\text{Cu}}=\SI{5.8e7}{\siemens\per\meter}$ giving $R_{\rm{s}}=\SI{29}{\milli\ohm}$ at \SI{10}{\giga\hertz}. In this design, $B_{\rm{vac}}^{\rm{rms}}=\SI{15.0}{\pico\tesla}$, which is greatly enhanced compared to a simple hollow three-dimensional cavity, with $B_{\rm{vac}}^{\rm{rms}}$ of typically \SI{1}{\pico\tesla}, or \SI{5}{\pico\tesla} for a very flat cylindrical cavity with losses similar to this LGR. This enhancement is a characteristic of the three-dimensional lumped-element geometry \cite{Angerer2016,Ball2018}. 

An important aspect in the design of the LGR is that several magnetic insulators of interest, such as iron garnets, are grown on substrates with a significant relative dielectric permittivity, reaching for instance $\varepsilon_{\rm{r}}=12$ for Gd$_{3}$Ga$_{5}$O$_{12}$ (GGG) or $\varepsilon_{\rm{r}}=13$ for Y$_{3}$Sc$_{2}$Ga$_{3}$O$_{12}$ (YSGG) \cite{Shannon1990}. Consequently, the capacitance across both faces of the sample loop, separated by \edit{$t$}, is not entirely negligible compared to the capacitance of the air-filled gaps with thickness \edit{$t'$}, which is suitably accounted for by the finite-element simulation. Beyond shifting down the operation frequency $\omega_0/(2\pi)$, this also slightly affects the uniformity of the magnetic component of the electromagnetic field, $\sigma_{\rm{B}}/B$, as shown in Table~\ref{tab:sim_results}.

A scale-independent metric for the magnetic component of the electromagnetic field, determined by structural design, is given by the rms value at the maximum of field $B_{\rm{G}}=B_{\rm{vac}}^{\rm{max,rms}}/f_0^2$, \edit{which corrects for the influence of the cavity frequency $f_0=\omega_0/(2\pi)$}, while a material-independent
metric for the losses of a resonator is $G=QR_{\rm{s}}$ \cite{Goryachev2014,Choi2023}. \edit{The quantity $G$ is a geometric factor that quantifies the spread of the spatial distribution of surface current density in the resonator. It decreases due to current crowding, when the magnetic component of the electromagnetic field is further concentrated in space.} For the present unloaded LGR, we find $B_{\rm{G}}=\SI{0.13}{\pico\tesla\per\giga\hertz\squared}$ and $G=\SI{23.4}{\ohm}$ (without sample) or $B_{\rm{G}}=\SI{0.22}{\pico\tesla\per\giga\hertz\squared}$ and $G=\SI{20.6}{\ohm}$ (with YSGG substrate). Ideally, both $B_{\rm{G}}$ and $G$ should be maximized, but a trade-off between strong current densities for strong $B_{\rm{vac}}^{\rm{rms}}$ and strong losses implies a balance between these two values \cite{Choi2023}. The present LGR design offers a combination of these two parameters that is very suitable for thin films, since the magnetic mode volume is almost that of the substrate, with a very good filling factor. The present LGR design features a geometrical factor $G$ that belongs to the low end of the range of values that are attainable for the corresponding value of $B_{\rm{G}}$ \cite{Choi2023}, because our design favors the homogeneity of $B_{\rm{vac}}^{\rm{rms}}$ over the whole sample space. \edit{For a typical \SI{200}{\nano\meter}-thick film of magnetic insulator YIG placed in the sample space ($N=\num{1e17}$), a significant coupling $g/(2\pi)\approx\SI{60}{\mega\hertz}$ is expected.} We note that if instead of a film, a prismatic bulk YIG sample was employed ($N=\num{2.5e20}$), the figures would reach $g/(2\pi)\approx\SI{2.9}{\giga\hertz}$, that is, $1/3$ of $\omega_0/(2\pi)$, and $\mathcal{C}\approx\num{1.5e6}$ at room temperature, or $\mathcal{C}>\num{5e6}$ at \SI{1}{\kelvin} (considering $\kappa_{\rm{m}}/(2\pi)\approx\SI{2}{\mega\hertz}$ in each case), which compares very well with other designs \cite{Goryachev2014,Bourhill2016}.

\footnotesize
\begin{table*}
    \centering
    \caption{Characteristic parameters for the LGR2 design with different sizes of the sample loop and gap thickness, using the FEM model, without sample. Columns are the dimensions of the sample loop $l$ and $d$; gap thickness \edit{$t'$}; resonance frequency; quality factor without sample nor external coupling from antennas; coupling magnetic field expressed as $B_{\rm{vac}}^{\rm{max}}$, its value at the LGR center; corresponding $B_{\rm{vac}}^{\rm{rms}}$ averaged over the sample volume; resonance linewidth; filling factors $\eta$ of the sample loop of $l\times{}d\times0.55$~\si{\milli\metre} and $\eta_{\rm{S}}$ of a sample scaled by the same factor within the magnetic mode volume; and standard deviation of $B_{\rm{vac}}^{\rm{rms}}$ in the scaled sample volume.}
    \label{tab:var_results}
    
    \begin{tabular*}{\textwidth}{>{\centering}p{1.52cm}>{\centering}p{1.52cm}>{\centering}p{1.52cm}@{\extracolsep{\fill}}cccccccc}
        \hline\hline
         $l$ & $d$ & \edit{$t'$} & $\omega_{\rm{0}}/(2\pi)$ 
         & \multirow{2}{*}{$Q^\mathrm{unloaded}$} 
         & $B_{\rm{vac}}^{\rm{max}}$
         & $B_{\rm{vac}}^{\rm{rms}}$
         & $\kappa_{\rm{c}}/(2\pi)$
         & \multirow{2}{*}{$\eta$}
         & \multirow{2}{*}{$\eta_\mathrm{S}$}
         & \multirow{2}{*}{$\sigma_{\rm{B}}/B$} \\
         (mm) & (mm) & (mm) & (GHz) & & (pT) & (pT) & (MHz) & & \\
        \hline
        5.1 & 5.5 & 0.1 & 11.08 & 849 & 22.6 & 15.0 & 13.1 & 0.754 & 0.621 & 0.055 \\
        5.1 & 2.75 & 0.05 & 11.88 & 694 & 32.9 & 22.7 & 17.1 & 0.804 & 0.652 & 0.029 \\
        5.1 & 11 & 0.2 & 9.21 & 927 & 15.6 & 9.1 & 9.9 & 0.695 & 0.594 & 0.175 \\
        2.55 & 5.5 & 0.1 &11.58 & 917 & 29.5 & 19.7 & 12.6 & 0.626 & 0.512 & 0.057 \\
        10.2 & 5.5 & 0.1 & 10.75 & 813 & 16.7 & 11.0 & 13.2 & 0.838 & 0.692 & 0.056 \\
        25.58 & 5.5 & 0.1 & 10.6 & 796 & 11.1 & 7.1 & 13.3 & 0.898 & 0.742 & 0.065 \\
        5 $\times$ 5.1 & 5.5 & 0.1 & 10.6 & 789 & 11.1 & 7.1 & 13.4 & 0.898 & 0.744 & 0.063 \\
        \hline\hline
    \end{tabular*}
\end{table*}
\normalsize

\subsection{Modularity}
\begin{figure}[t]
    \centering
    \includegraphics[width=3.5in]{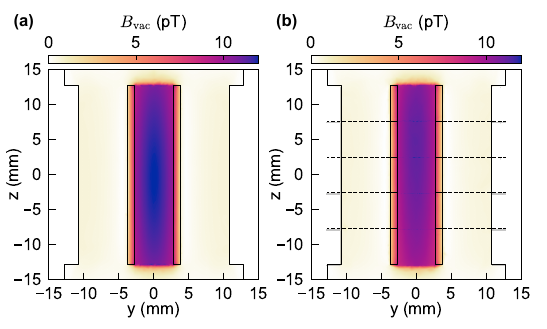}
    \caption{Intensity maps of the coupling field $B\textsubscript{vac}$ in the sample plane ($yz$), as determined by finite-element simulations, for \panel{a}an LGR same as in Fig.~\ref{fig:01}d, but $l=\SI{25.58}{\milli\meter}$, \panel{b}an LGR made of 5 identical modules with $l=\SI{5.1}{\milli\meter}$, each separated by \SI{0.02}{\milli\meter}. The black rectangles locate the two gaps, the dashed lines separate the different LGR modules.}
    \label{fig:02}
\end{figure}

The two lids of the cavity box and the two halves of the LGR can be modified and reassembled independently. This modular approach allows for modifications of the design according to the specifics of a sample. In particular, different frequencies can be probed by assembling LGRs with different gaps. To illustrate the modularity of the present LGR geometry, we present in Fig.~\ref{fig:02} one possible variation of our resonator design based on using several loop-gap modules identical to the one presented earlier, assembled by aligning their loops along $z$, and enclosed by the two same cylindrical lids shown in Fig.~\ref{fig:01}a. The several loop-gap modules can even be separated by a small offset (for instance here, \SI{0.02}{\milli\meter}), as no currents flow along $z$. The direct comparison between Figs.~\ref{fig:02}a and \ref{fig:02}b demonstrate that the LGR made of 5 identical modules behaves almost identically to an LGR with 5 times larger $l$. This approach enables a large volume of homogeneous microwave excitation for applications where a large volume of magnetic material is required \cite{Flower2019a,Crescini2020b}.

\begin{figure}[t]
    \centering
    \includegraphics[width=3.5in]{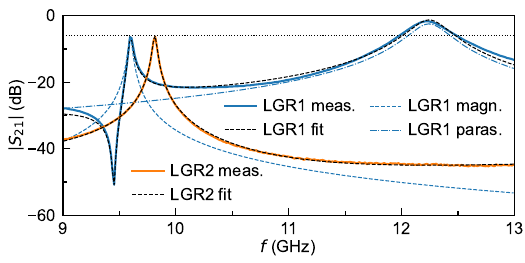}
    \caption{\edit{Microwave transmission $\norm{S_{21}}$ for both LGR1 and LGR2. Blue and orange solid lines correspond to the measured $\norm{S_{21}}$ for the loaded LGR1 and LGR2, respectively, without magnetic field applied. The horizontal dotted line at \SI{-6}{dB} represents the transmission at critical coupling conditions. Black dashed lines correspond to the fitted cavity transmission for LGR1 and LGR2. For LGR1, the fit line is decomposed into the contributions from the magnetically coupled main LGR mode (LGR1 magn.) and from the parasitic mode near \SI{12}{\giga\hertz} (LGR1 paras.), whose complex sum provides the fitted $S_{21}$.}}
    \label{fig:03}
\end{figure}

We have also modified the design of the upper cavity box to eliminate a parasitic mode near the main mode of interest (more details in Appendix~\ref{app:optim}). This is achieved by inserting a separator plate between the antennas. This modification results in two designs of LGR that we employ in the following: LGR1 (no plate), which has a parasitic mode slightly above \SI{12}{\giga\hertz}, near the mode of interest around \SI{10}{\giga\hertz}; and LGR2 (with separator plate), which features an undisturbed transmission around the main resonance at \SI{10}{\giga\hertz}. \edit{The microwave transmissions for LGR1 and LGR2 are shown in Fig.~\ref{fig:03}, which also presents fits to the interfering microwave transmission from the main mode of the LGR (magnetically coupled to the sample space) and the nearby parasitic mode (box mode). These fits to the sum of complex-valued transmission from neighboring modes enable reliable extraction of the resonance linewidth of the main LGR mode, even when a Fano-shaped resonance is observed.}

\subsection{Scaling laws}
From the circuit model approximation, we can briefly summarize the most relevant scaling laws for adapting the design to a specific sample: (i) $\omega_0/(2\pi)$ is roughly independent from $l$ and is tuned by \edit{$\sqrt{t'/d}$}. Although the actual current distributions will evolve compared to this simple lumped-element model, $\omega_0/(2\pi)$ should also be roughly proportional to \edit{$1/\sqrt{d't}$}; (ii) $Q$ evolves as \edit{$\omega_0^{1/2}(dt)/(d+t+d')$}; (iii) $B_{\rm{vac}}^{\rm{rms}}$ scales as \edit{$\sqrt{\omega_0/(ldt)}$}. Since $g\propto\sqrt{N}B_{\rm{vac}}^{\rm{rms}}$, $g$ is independent from $V$ or $V_{\rm{m}}$ if this volume is filled with the substrate at constant filling factor \cite{Huebl2013}. \edit{These scaling laws are illustrated for LGR2 in Table~\ref{tab:var_results}, and are overall verified if we compare line-by-line the seven sets of LGR dimensions tested here by finite-element modeling.} The actual current distributions in the different elements of the LGR explain the small deviations away from these scaling laws. For instance, reducing $d$ and \edit{$t'$} by a common factor roughly maintain $\omega_0$, while it improves the homogeneity of the excitation at the cost of a slightly higher cavity dissipation.

Developing the terms entering in the cooperativity, we thus find
\begin{equation}
\label{eq:C_long}
\edit{
    \mathcal{C}=\frac{\mu_0^{3/2}\gamma^2\hbar}{2\sqrt{2}}\sqrt{\sigma_{\rm{Cu}}}n_{\rm{s}}\eta_\mathrm{S}t_{\rm{F}}\frac{d}{d+t+d'}\frac{Q_{\rm{m}}}{\sqrt{\omega_0}}
}
\end{equation}
with $t_{\rm{F}}$ the magnetic film thickness, $\eta_\mathrm{S}$ the filling factor of the whole sample (substrate+film) in the LGR magnetic mode, and $Q_{\rm{m}}=\omega_0/\kappa_{\rm{m}}$ the magnetic quality factor. The geometric factor \edit{$d/(d+t+d')$} is in any case close to unity, so that for a given $t_{\rm{F}}$ and magnetic material spin number density $n_{\rm{s}}$, $\mathcal{C}$ only depends significantly on the frequency chosen for the LGR. The magnetically ordered system has a resonance linewidth that usually follows $\kappa_{\rm{m}}=\kappa_{\rm{inh}}+2\alpha\omega_0$, with $\kappa_{\rm{inh}}$ the inhomogenous contribution to the linewidth and $\alpha$ the dimensionless Gilbert damping. Hence, the last term in Eq.~\eqref{eq:C_long}, which can be written $\sqrt{\omega_0}/\kappa_{\rm{m}}$, is maximized at a finite value of $\omega_0=\kappa_{\rm{inh}}/(2\alpha)$. 

\section{Strong coupling to thin film YIG}

\begin{figure}[t]
    \centering
    \includegraphics[width=3.5in]{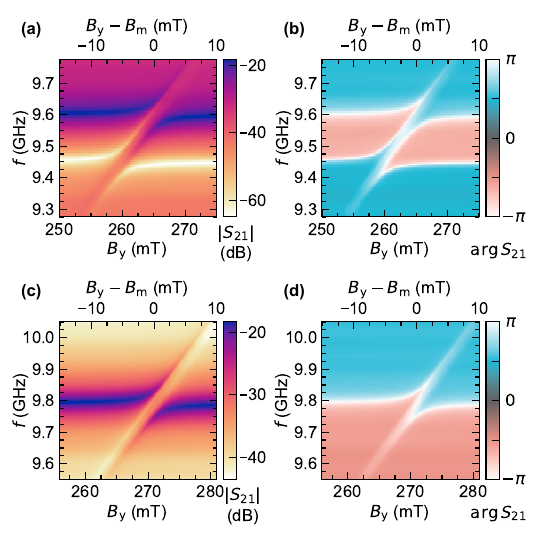}
    \caption{Microwave transmission of the loaded LGRs with \SI{75}{\nano\meter}-thick YIG film. \panel{a}Amplitude and \panel{b}phase of $S_{21}$ as a function of in-plane field $B_{\rm{y}}$ and measurement frequency $f$, for LGR1 (hollow upper cylindrical box). \panel{c}Amplitude and \panel{d}phase of $S_{21}$ as a function of $B_{\rm{y}}$ and $f$, for LGR2 (cylindrical box with separator). $B_{\rm{m}}$ is the field required to excite the Kittel mode of the film at the cavity resonant frequency.}
    \label{fig:04}
\end{figure}

\begin{figure}[t]
    \centering
    \includegraphics[width=3.5in]{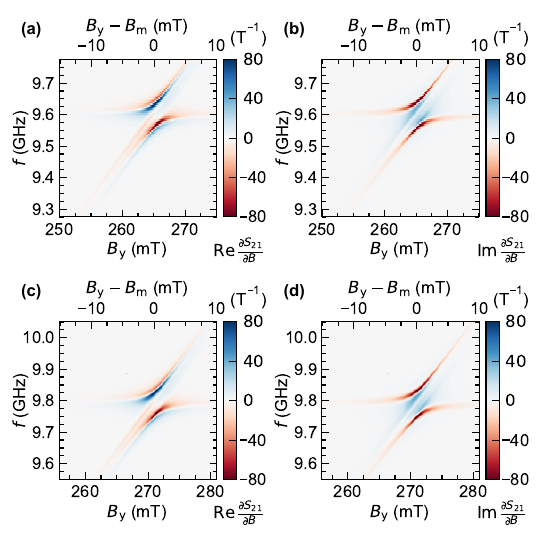}
    \caption{Field-derivative microwave transmission of the loaded LGRs with \SI{75}{\nano\meter}-thick YIG film. \panel{a}Real and \panel{b}imaginary parts of $\partial{}S_{21}/\partial{}B$ as a function of in-plane field $B_{\rm{y}}$ and measurement frequency $f$, for LGR1 (hollow upper cylindrical box). \panel{c}Real and \panel{d}imaginary parts of $\partial{}S_{21}/\partial{}B$ as a function of $B_{\rm{y}}$ and $f$, for LGR2 (cylindrical box with separator). The sinusoidal modulation field is \SI{40}{\micro\tesla}.}
    \label{fig:05}
\end{figure}

We describe here the measurement of the LGRs loaded with a \SI{75}{\nano\meter}-thick epitaxial YIG film. The film was grown by magnetron sputtering \cite{Legrand2025a} on a \qtyproduct{5x5x0.5}{\milli\meter} substrate of YSGG. All the measurements that follow are performed at room temperature. The present film is characterized by an inhomogeneous linewidth $\Delta{}B_0=$~\SI{0.23}{\milli\tesla} and a Gilbert damping $\alpha=7.34\times10^{-4}$, corresponding to a broadband ferromagnetic resonance (FMR) linewidth $\Delta{}B=$~\SI{0.75}{\milli\tesla} at \SI{10}{\giga\hertz}. Figure~\ref{fig:04} shows the complex microwave transmission parameter $S_{21}$ for the two designs of LGR, with magnetic field applied in the film plane along $y$. In the first design (LGR1), a broad resonant mode at \SI{12}{\giga\hertz} \edit{(visible in Fig.~\ref{fig:03}, mode labeled (g) in Fig.~\ref{fig:07} of Appendix~\ref{app:optim})} causes a complex background adding to the transmission of the mode of interest, leading to destructive interference near \SI{9.45}{\giga\hertz} (Fano-shaped resonance). Although this destructive interference condition shifts with magnetic field, and thus resembles an anti-crossing, \edit{it is due to the transmission of a mode of the LGR system that is not magnetically coupled to the sample space}. The data can be fit with the transmission derived from input-output theory according to \cite{Harder2018, Clerk2010, Li2020}
\begin{equation}
\label{eq:S21}
\edit{
    S_{21}(\omega, B_0)=
    \frac{A\,\kappa_{\rm{c}}}{ 2i (\omega_{\rm{c}} - \omega) - \kappa_{\rm{c}} + \frac{2g^2 }{i (\omega_{\rm{m}} - \omega) - \kappa_{\rm{m}} } }+F,
    }
\end{equation}
where $A$ is a transmission ratio related to the ratio of internal to external Q factors of the LGR (or coupling regime), $\omega_{\rm{m}}=\gamma  \sqrt{B_{\rm{y}} (B_{\rm{y}}+\mu_0 M_{\rm{eff}})} $ is the Kittel formula giving the magnetic resonance frequency for static magnetic field $B_{\rm{y}}$ applied in the sample plane, with $\mu_0 M_{\rm{eff}}$ the effective magnetization, \edit{and $F$ is a slow-varying, complex background due to the transmission of a nearby parasitic mode, see Appendix~\ref{app:optim}}. The cooperativity resulting from the fit is $\mathcal{C}=7.4\pm0.5$, with $g/(2\pi)=$~\SI{39.3}{\mega\hertz}, larger than both $\kappa_{\rm{c}}/(2\pi)=$~\SI{32.1}{\mega\hertz} and  $\kappa_{\rm{m}}/(2\pi)=$~\SI{26.0}{\mega\hertz} and thus reaching the strong-coupling regime. The value of $g/(2\pi)$ extracted from the fit is consistent with the modeling performed above. \edit{A coupling $g/(2\pi)=\SI{40.8}{\mega\hertz}$ is predicted for the present YIG film with $N=\num{3.75e16}$ spins, which needs to be corrected for the reduced magnetization at room temperature, compared to bulk YIG ($M_{\rm{S}}=\SI{195}{\kilo\ampere\per\meter}$) in the zero-temperature limit. We use room-temperature SQUID magnetometry to directly find the effective magnetic moment, $N=\num{3e16}$, so an effective coupling $g/(2\pi)=\SI{36.5}{\mega\hertz}$ is expected.} In the modified design (LGR2), the presence of a separator plate in the $xz$ plane within the lid breaks the rotational symmetry in the upper half of the cavity box and eliminates the detrimental transverse magnetic mode. A usual anti-crossing behavior without background is then observed. \edit{The fits determine $g/(2\pi)=$~\SI{38.8}{\mega\hertz}, $\kappa_{\rm{c}}/(2\pi)=$~\SI{33.5}{\mega\hertz}, and  $\kappa_{\rm{m}}/(2\pi)=$~\SI{27.5}{\mega\hertz},} resulting in an estimated cooperativity very close to that of LGR1 with $\mathcal{C}=6.5\pm0.3$.

Measuring a field-derivative transmission by recording the variations of $S_{21}$ with respect to a sinusoidal modulation of the magnetic field is a common procedure to improve the signal-to-noise ratio of FMR measurements. We show here that this is a meaningful strategy to eliminate the parasitic modes of lumped-element magnon--photon systems, since those do not couple with the magnetic sample. \edit{The setup that we use for field-differential transmission measurements is based on the lock-in detection of microwave amplitude and phase variations with respect to modulation of the external static field, and is comprehensively described in our previous work \cite{Legrand2025}.} Figure~\ref{fig:05} shows the field-derivative data for LGR1 and LGR2, which both display a clean, single anti-crossing pattern without background interference. Whereas the bright mode is coupled to the magnon system and has a transmission that depends on external field $B_{\rm{y}}$, the parasitic dark mode of LGR1 \edit{entering as $F$ in Eq.~\eqref{eq:S21} has no field dependence, leading to $\partial{}F/\partial{}B=0$}. \edit{Although interference with this parasitic mode creates the Fano-shaped anti-resonance branches in the direct transmission data, this effect is thereby eliminated in the $\partial{}S_{21}/\partial{}B$ measurement.} Therefore, field-differential measurements are extremely useful to isolate the modes of interest in hybrid magnon--photon systems at room temperature. Although more difficult to implement at low temperatures due to Eddy current heating and the large inductance of superconducting magnets, a careful approach to field-modulation in cryogenic conditions would result in a very sensitive detection of hybridized modes in magnetic insulator thin films.

A fit with the expected complex-valued derivatives of $S_{21}$ results \edit{in similar values of magnon--photon coupling and magnon linewidth as above, $g/(2\pi)=$~\SI{41.2}{\mega\hertz} and  $\kappa_{\rm{m}}/(2\pi)=$~\SI{26.2}{\mega\hertz} for LGR1, $g/(2\pi)=$~\SI{41.0}{\mega\hertz} and  $\kappa_{\rm{m}}/(2\pi)=$~\SI{27.0}{\mega\hertz} for LGR2. Considering the values of cavity linewidth $\kappa_{\rm{c}}/(2\pi)$ previously obtained with better precision from the direct measurement of $S_{21}$,} the cooperativities are in line with above, $\mathcal{C}= 8.1 \pm 0.8$ for LGR1 and $\mathcal{C}=7.4 \pm 0.5$ for LGR2. \edit{A small deviation of $\mathcal{C}$, within the fit errors, is observed between the two methods. It originates in the non-ideal lineshape of the magnetic resonance, for which inhomogeneous broadening is present. The convolution between the ideal Lorentzian lineshape and the approximately Gaussian lineshape due to inhomogeneous broadening is impacting differently the direct transmission and the field-derivative fits \cite{Legrand2025}}.

In the present LGR geometry, the resulting magnon--photon coupling is twice as large as that of a YIG film 50 times as thick placed in a prismatic hollow cavity resonator \cite{Zhang2016e}. Some additional lines appear within the anti-crossing gap of Fig.~\ref{fig:04}, and more clearly in Fig.~\ref{fig:05}. These lines are assigned to other magnon modes than the homogeneous FMR mode \cite{Zhang2016e}, structured by the dipolar interactions within the magnon system, but with much less coupling to the cavity mode of interest \cite{Goryachev2014}.

\section{Coupling to standing spin-wave modes}

\begin{figure}[t]
    \centering
    \includegraphics[width=3.5in]{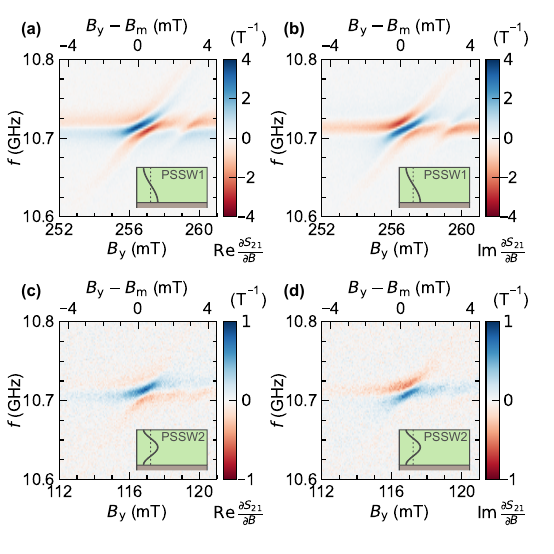}
    \caption{Field-derivative microwave transmission for PSSW modes in the LGR2 system with \SI{75}{\nano\meter}-thick YIG film. \panel{a}Real and \panel{b}imaginary parts of $\partial{}S_{21}/\partial{}B$ as a function of in-plane field $B_{\rm{y}}$ and measurement frequency $f$, for PSSW mode number 1. \panel{c}Real and \panel{d}imaginary parts of $\partial{}S_{21}/\partial{}B$ as a function of $B_{\rm{y}}$ and $f$, for PSSW mode number 2. The sinusoidal modulation field is \SI{40}{\micro\tesla}.}
    \label{fig:06}
\end{figure}

Resorting to field-differential measurements improves the sensitivity to small signals. In particular, we demonstrate that our approach enables the detection of magnon modes having a finite and quantized wave-vector across the thickness of the film, corresponding to perpendicular standing spin wave (PSSW) modes \cite{Klingler2015}. Except in the presence of surface pinning \cite{Cao2015}, these modes have a vanishing spatial overlap with a homogeneous magnetic field excitation or probing field across a scale of 10--\SI{100}{\nano\meter}, the thickness of the magnon system. In the present high-quality YIG sample, the PSSW modes are very weakly pinned, and are therefore barely coupled to the probing field. This results in a measurement that does not exhibit a clear gap between the magnon--photon polaritonic branches, and considerably less contrast compared to the FMR mode, about 20 times weaker for the first PSSW mode and 80 times weaker for the second PSSW mode. \edit{For this measurement, a slightly modified version of the LGR system is used, described as LGR3 in Appendix~\ref{app:optim}. Hence, the cavity resonance occurs here at about \SI{10.7}{\giga\hertz}, slightly different from LGR1 and LGR2.} Despite the weak coupling of macroscopic resonators to these modes, they can still be detected in the LGR, as shown in Fig.~\ref{fig:06}. \edit{The first two PSSW modes with $k_{\rm{z}}=\pi/t_{\rm{F}}$ and $k_{\rm{z}}=2\pi/t_{\rm{F}}$ can be observed in Figs.~\ref{fig:06}a,b and Figs.~\ref{fig:06}c,d, respectively.} Similarly to the main FMR mode, an additional line is visible on the right of the main PSSW1 mode in Figs.~\ref{fig:06}a and b, which corresponds to a $k_{\rm{z}}=\pi/t_{\rm{F}}$ but laterally inhomogeneous mode.

\section{Conclusions}

In conclusion, we have designed a three-dimensional LGR which maximizes the filling factor of magnon--photon hybrid systems based on magnetic insulator thin films. We describe the LGR optimization procedure with its main characteristics and scaling laws. We experimentally demonstrate that this LGR reaches the strong-coupling regime with a \SI{75}{\nano\meter}-thick sputtered film of YIG. We also show how to isolate the modes of interest by field-differential magneto-spectroscopy, eliminating the signal from the uncoupled dark modes. Using this approach, we can access the weakly coupled perpendicular standing spin-wave modes arising in thin films. The LGR magnon--photon coupling geometry is particularly suitable for ultrathin films of magnetic insulators, as it enables the strong-coupling regime while maintaining a sufficient homogeneity of the magnetic component of the electromagnetic field. This geometry notably facilitates the investigation of a wide spectrum of magnon--photon hybridization phenomena, with tunable magnon properties across cryogenic to room temperature and above, all with a single modular resonator system. 

In the future, hybrid magnon--photon cavity systems are envisioned for use in quantum information \cite{Soykal2010,Tabuchi2014,Tabuchi2015}, with the perspective of enabling a tunable coherent conversion between microwave and optical photons \cite{Hisatomi2016,Engelhardt2022,Rameshti2022}, high-sensitivity magnetometry \cite{Flower2019a,Crescini2020b}, investigation of magnon-based quantum states \cite{Kamra2020,Sharma2021,Sun2021}, and mediation of entanglement between spatially separated spin qubits \cite{Fukami2021}. Making progress in the isolation of the modes of interest, in the detection of exchange magnon modes and in the optimization of the cooperativity are essential to these objectives. Because the strong coupling of microwave photons to thin-film magnons enables a fundamentally different and tunable magnon mode structure in comparison to bulk spheres, we anticipate that our LGR design can optimally serve this purpose. Future work may focus on the cryogenic implementation of LGRs, which would benefit from better quality factors ($R_{\rm{s}}\approx\SI{10}{\milli\ohm}$ or lower for OFHC copper at \SI{1}{\kelvin}, and much lower with superconducting coatings), combined with lower magnetic losses \cite{Guo2023,Legrand2025a}, which promise very high cooperativities at low temperatures.

%

\section*{Acknowledgments}

We acknowledge F.~Nasr for help with SQUID magnetometry. F.Z.~was supported by the Chair ``Innovative Processes and Materials" led by l'X-{\'{E}}cole Polytechnique and the Fondation de l’{\'{E}}cole Polytechnique, sponsored by Saint-Gobain. H.W.~acknowledges the support of the China Scholarship Council (CSC, Grant No.\ 202206020091). W.L.~acknowledges the support of the ETH Zurich Postdoctoral Fellowship Program (21-1 FEL-48). 

\appendix

\section{Details on the loop-gap resonator optimization}
\label{app:optim}

\begin{figure*}[ht]
    \centering
    \includegraphics[width=\textwidth]{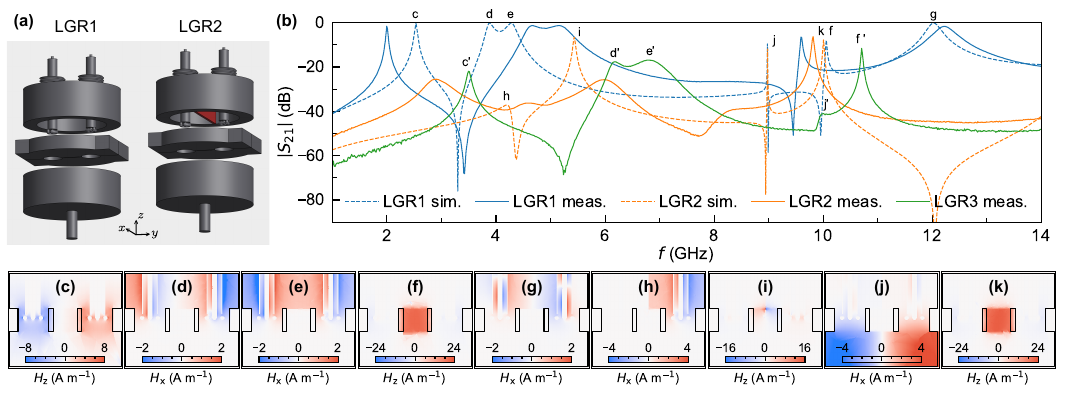}
    \caption{Microwave transmission $\norm{S_{21}}$ and modes in LGR1/LGR3 and LGR2. \panel{a}Views of the modeled resonators, \panel{b}simulated and measured $\norm{S_{21}}$ for LGR1 and LGR2; measured $\norm{S_{21}}$ for LGR3. \panel{c--k}Corresponding maps of $H_{\rm{x,z}}$ excitation field components for each resonance mode labeled in panel b, with a \SI{0}{dBm} input power. Panels c--g are LGR1 modes, h--k are LGR2 modes. Modes c--f and j are found again as modes c'--f' and j' for LGR3. The field of view is enlarged compared to that in Fig.~\ref{fig:01}d, the external boundary lines in black are the metallic walls.}
    \label{fig:07}
\end{figure*}

In this Appendix section, we detail the optimization of the well-known LGR geometry for a use with thin films of magnetic insulators. For a defined sample shape, here a magnetic insulator thin film grown on a \qtyproduct[product-units=power]{5 x 5 x 0.5}{\milli\metre} prismatic substrate, the objectives for the design optimization are: achieving a resonance frequency of the loaded LGR $\omega_0/(2\pi)\approx$~\SI{10}{\giga\hertz} (typical for FMR of iron garnets obtained by physical vapor deposition techniques, as a trade-off between an FMR linewidth dominated by inhomogeneous broadening, and too much increase in linewidth due to Gilbert damping at large frequencies), obtaining a critical coupling with external transmission lines (so that a large signal contrast can be reached at room temperature, while the external losses can always be reduced by retracting the probes), maximizing the filling factor, and maintaining sufficient $B$-field homogeneity. 

The electric component of the electromagnetic field is confined near the gaps, across which opposite surface charge densities accumulate. Conversely, surface currents flow mainly around the sample loop, in which the magnetic field is concentrated. The shape of the sample loop is the main critical parameter defining the magnetic coupling, where a square section is preferable to a cylindrical one for a thin-film system on a substrate. Having $d$ and $t$ only a fraction of \si{\milli\meter} larger than the sample concentrates the effective magnetic mode volume. A surface current of much lower magnitude and opposite direction flows in the return-flux loops. Dissipation essentially originates from the surface currents in the sample loop. The dimensions and shape of the return-flux loops have a minor effect on the mode shape and frequency, as long as their cross-sectional surface is large enough for their self-inductance to be negligible compared to that of the sample loop.

With the main self-inductance $L$ fixed by the sample loop, $\omega_0$ is then primarily controlled by the capacitance $C$, which in turn depends on the number of gaps and their dimensions. Increasing the number of gaps in the loop-gap module to four or more \cite{Webb2014, Eisenach2018, Tschaggelar2017} could be pursued, but would increase the overall complexity and lead to additional unwanted loop-gap modes. Varying the thickness and width of the two gaps is a more direct approach, as it only slightly modifies the shape of the mode of interest while setting $\omega_0$. The design made of two halves enables precision machining of the gap $t'$, by adjusting the height offset between the gap plates and the contacted surfaces of the LGR. Here, the accurate control of $t'$ from \SIrange{50}{150}{\micro\meter} enables reaching $\omega_0/(2\pi)\approx$~\SI{10}{\giga\hertz}.

Two common approaches to couple the LGR to external signals are capacitive, by approaching a pin-shaped antenna near the gaps, and inductive, by inserting a loop-shaped antenna near the sample or return-flux loops \cite{Rinard1993, Wood1984}. Pin-shaped antennas can be made from the central conductor of a coaxial cable, forming a three-plate capacitor with the gap. Loop-shaped antennas can be made by bending the inner conductor of a coaxial cable and shorting it to the outer conductor, forming coupled solenoids with the nearby loop in which magnetic fields are coupled by mutual inductance \cite{Ghamsari2013}. We chose the latter due to its easier implementation of suitable external couplings, and to avoid the changes of capacitive coupling due to variations of $\varepsilon_{\rm{r}}$ of each sample, as unavoidably, the capacitance partially leaks through the sample space. Since $\omega_0$ can be finely adjusted with $t'$, while the geometry of the loop elements is fixed by the sample dimensions, it is convenient that the inductive coupling to the loop-shaped antennas can be tuned almost independently of the LGR itself. To achieve critical coupling \cite{Pozar2012} with loop-shaped antennas, their loop axis can be aligned with the return-flux loops, in a region where the $B$-field is sufficiently strong. It is chosen here to insert symmetric antennas inside each return-flux loop, fine-tuning their position along $z$ using a screw system (see Fig.~\ref{fig:01}a). To maximize the inductive coupling and to allow for the widest tunability from under-coupled to over-coupled regimes, the diameters of the return-flux loops and of the antennas should differ only by a few \si{\milli\meter}. The antennas could also be placed at the two opposite sides of the sample loop, where the magnitude of the magnetic field is considerably high, but this would make the system less easy to integrate. The positioning of the input and output antennas on only one side of the $xy$ sample plane facilitates the mounting and rotation of the resonator from the other side.

In a last step, the LGR is enclosed between two halves of a cylindrical cavity box, which can be rotated around its axis to perform angular-dependent measurements. The rough optimization based on the circuit model is finally refined with a finite-element electromagnetic solver. The simulated properties for LGR1 in Table \ref{tab:sim_results} are obtained after optimizing the system dimensions according to the steps provided above. The inductive couplings with the exact antenna shapes are then introduced in the solver to simulate the resonator transmission across a broad range of excitation frequencies. All resonance modes of the system up to \SI{14}{\giga\hertz}, including undesired modes that are inefficient for magnetic coupling, are characterized by a Vector Network Analyzer broadband measurement, and compared to the finite-element predictions, as shown in Fig.~\ref{fig:07}. Field maps are displayed for each mode in Figs.~\ref{fig:07}c--g. The mode at \SI{2}{\giga\hertz} is a loop-gap mode where no magnetic field circulates in the sample loop, and the parasitic modes at 3.8--\SI{4.2}{\giga\hertz} and \SI{12}{\giga\hertz} correspond to excitations of the upper half of the cavity box where the magnetic field is transverse and rotates around the $z$-axis. 

The microwave transmission of LGR1 in the frequency range of interest for this work can be fitted as the sum of transmissions from  modes (f) and (g). To include the box mode (g), Eq.~\eqref{eq:S21} contains $F$, given by
\begin{equation}
\label{eq:S21_F}
\edit{
    F=
    \frac{A'\,\kappa_{\rm{f}}}{ 2i (\omega_{\rm{f}} - \omega) - \kappa_{\rm{f}}},
    }
\end{equation}
with $A'$ a transmission factor related to the ratio of internal to external Q factors of the box mode, $\kappa_{\rm{f}}/(2\pi)$ the linewidth of the box mode, and $\omega_{\rm{f}}/(2\pi)$ the frequency of the box mode.

The two lids of the cavity box and the two halves of the LGR can be modified and re-assembled independently. We have for instance modified the design of the upper cavity box to eliminate the parasitic mode at \SI{12}{\giga\hertz} \edit{labeled (g) for LGR1}, simply inserting a separator plate (LGR2), as demonstrated in Fig.~\ref{fig:07}. The respective field maps are displayed for each mode in Figs.~\ref{fig:07}h--k.

\edit{The measurements of the coupling to PSSW modes were performed with another modified version of LGR1, which had a hollow box of smaller diameter, and a slightly increased gap spacing $t'$. The transmission of this system is displayed as LGR3 in Fig.~\ref{fig:07}b. Due to the reduction of diameter of the surrounding box, modes (c), (d), (e) and (j) are shifted to higher frequencies, and their peak transmission is reduced. The new positions for these modes are identified as (c'), (d'), (e') and (j') in Fig.~\ref{fig:07}b. The further adjusted gap $t'$ also causes a shift of the main mode (f) to a higher frequency around \SI{10.7}{\giga\hertz}, here labeled (f') for LGR3. This prevents the overlap between modes (j') and (f'). The previous box mode (g) is pushed far from (f'), and thus it does not cause any visible Fano-shaped resonance around mode (f').}

\bibliography{Loop_gap}

\end{document}